\documentclass[preprint,showpacs,preprintnumbers,amsmath,amssymb]{revtex4}
\usepackage{graphicx}% Include figure files
\usepackage{dcolumn}% Align table columns on decimal point
\usepackage{bm}% bold math

\begin{document}
\title{First-passage probability: a test for DNA Hamiltonian parameters}

\author{Marco Zoli}

\affiliation{School of Science and Technology \\  University of Camerino, I-62032 Camerino, Italy \\ marco.zoli@unicam.it}

\date{\today}

\begin{abstract}
A method is proposed to select the suitable sets of potential parameters for a one-dimensional mesoscopic Hamiltonian model, first introduced to describe the DNA melting transition and later extended to investigate thermodynamic and dynamical properties of nucleic acids. The DNA base pair fluctuations are considered as time dependent trajectories whose initial condition sets the no crossing constraint enforced in the path integral for the first-passage probability. Performing the path integration at room temperature, relations are established between the cutoff on the amplitude of the base pair fluctuations and the model parameters. In particular, it is shown that the non-linear stacking parameter should be $\sim 1$. The formalism here developed may be applied to compute the lifetime of open base pairs in three-dimensional helical models for DNA molecules.
\end{abstract}


\maketitle

\section*{I. Introduction}

While the DNA structure is stable at room temperature mostly due to the covalent bonds between adjacent nucleotides along the sugar-phosphate backbone, the fundamental biological processes of replication, transcription and protein binding rely on the local unzipping of the double helix  which allows for reading and
copying of the genetic code \cite{croq00,marcus13,biton18,lee19}. Thermal fluctuations can locally disrupt the DNA molecule, starting in regions rich in the weaker
$AT$ base pairs, and lead to transient formation of breathing bubbles  as the energy scale for the separation of the bonds between paired bases is $\sim 2 - 3 k_BT$  \cite{heslot97,bres86,ratner95}. 
While denaturation bubbles generally appear both in linear and circular supercoiled DNA as a response to release the torsional stress \cite{benham99}, their size and number varies with the ambient conditions, sequence heterogeneity and  chain length \cite{metz10}.  The denaturation transition of DNA in solution has been widely investigated mainly by monitoring the increase of UV absorbance ($\sim 260nm$) versus temperature due to the breaking of the hydrogen bonds and unstacking of planar adjacent bases. The melting temperature, usually defined by the mid-point transition at which half of the base pairs are broken, provides a measure of the relative content of $GC$ and $AT$ pairs in the sequence \cite{doty62}.

In labs, thermally driven separation of complementary strands is routinely carried out for several applications e.g., amplification of specific DNA fragments by polymerase-chain reaction \cite{mullis85} and genotyping \cite{zhou07}, genomic melting maps \cite{hovig07} and mutations detection in DNA microarrays \cite{bots08}, classification of DNA sequences and genetic distance between species \cite{wilson75}.

Besides its importance to molecular biology and biochemistry, the DNA melting, being a paradigm of a phase transition, has been the subject of a vast number of studies produced by the statistical physicists community over the last decades \cite{pol66,fish66,azb79,wart85}.
Computational methods based on Transfer Integrals  \cite{rapti06,singh11,hando12}, Monte Carlo simulations \cite{hwa05,ares05,bish09,olson10,kalos11}, molecular dynamics \cite{maiti15,tan17,skepo17,maiti17,lak19} and path integrals \cite{io13,io16,io18b}  have been also applied to coarse grained Hamiltonian and polymer physics models, such as the Worm-Like-Chain, to characterize the helix-coil transition, denaturation bubbles formation, force-extension behavior and flexibility properties of DNA helices e.g., persistence length and cyclization probability. While the WLC model has been known for long to provide an accurate description of long DNA sequences \cite{shimada84}, its effectiveness has been recently questioned at length scales smaller than the standard persistence length \cite{fenn08,archer08,gole12}. In fact, for short sequences, the enhanced bending flexibility associated with bubble formation \cite{marko04} may be responsible for the observed $J$-factors which appear larger than those predicted by the conventional WLC model \cite{vafa12,kim14}. Instead, at such scales, statistical mechanics methods based on mesoscopic Hamiltonian models suitably account for the bending and looping properties \cite{sung15,io16b}.  

A paradigm among the Hamiltonian approaches is the Peyrard-Bishop (PB) model \cite{pey89} which provides an appealing description of the double stranded chain in terms of the main forces acting at the level of the base pair i.e., hydrogen bonds between inter-strand pair mates modeled by a Morse potential and intra-strand harmonic stacking between adjacent bases. As the PB Hamiltonian is written in terms of a single degree of freedom, the relative distance between pair mates, the model is essentially one-dimensional and maps onto an exactly solvable Schr\"{o}dinger equation for a particle in a Morse potential which yields a crossover temperature characteristic of a smooth thermal denaturation. A later version of the PB model  incorporating nonlinear stacking interactions (also termed DPB model \cite{pey93}) has predicted a sharper melting transition although the real character of the denaturation may largely depend on the sequence specificities \cite{io10}.

While the simple representation for the helix, as made of two parallel strands, put forward by the PB model strictly holds in the denaturation regime, 
a more complex and realistic Hamiltonian model is now available to calculate thermodynamic and flexibility properties of short chains \cite{io14b,io17,io19a}. This 3D helical model goes beyond the PB picture, accounting for the twisting and bending of the chain and also for the presence of the solvent surrounding the DNA molecule. 

However, the quantitative predictions of the helical Hamiltonian model rely on a set of input potential parameters and this points to the importance of selecting appropriate sets for specific sequences.
Despite the considerable number of papers produced over the last decades with focus on the statistical mechanics of the PB Hamiltonian for DNA, no consensus has been found regarding the optimal data-set to be used in the models and significant variations are found as for the choice of the parameters \cite{zdrav06}.
For instance, careful estimates for the Morse potential depth and width and for the stacking force constant in the PB harmonic model have been obtained \cite{weber09a} by fitting the experimental melting temperatures for a set of short sequences, ranging between 10 and 30 base pairs, at various salt concentrations \cite{owcz04,lucia04}. Such values differ however, mostly as for the spatial range of the Morse potential, from the set of parameters obtained by fitting the melting temperatures of long homogeneous DNA chains, albeit via a PB anharmonic model \cite{campa98}. This latter set, in turn, is similar to the original values used in ref.\cite{pey93} except for the choice of the anharmonic stacking parameter ($\rho=\,2$) which is taken a factor four larger than in the 1993 work. Further, $\rho=\,25$ has been adopted in the analysis of the melting of short sequences \cite{pey09}  whereas later calculations of melting profiles of plasmid pBR322 and T7 phage \cite{the10}  and recent research on the melting of short sequences in crowded environments \cite{singh17} have even assumed $\rho=\,50$. 
Although such high anharmonic values may be consistent with a strong, temperature driven,  reduction of the persistence length in the transition to the single-stranded configuration, it remains that they have been  applied to model also double-stranded sequences well below the melting transition. Generally, it appears that $\rho$ is growing in the literature as a function of time and the noticed discrepancies are largely due to attempts to fit various experimental quantities by the same simple model and to the lack of precise constraints. 

These issues are addressed by the present investigation which aims to find a physical constraint for the selection of the model parameters. 
The strategy is the following.
Given that DNA thermodynamic properties are obtained by performing multiple integrals over the degrees of freedom for the nucleotides in the sequence, we notice that the computation always implies a somewhat arbitrary truncation of the phase space available to base pairs fluctuations. For instance,  transfer integral techniques for the PB  model require an upper cutoff in the solution of the kernel integral equation to avoid the divergence of the partition function \cite{zhang97}. While this cutoff should correspond to the largest allowed separation for the fluctuating base pair, large uncertainties exist as for the choice of the specific value.
I am suggesting here that, for a given sequence, the set of model parameters and in particular the non-linear stacking  should be physically related to the amplitude of the transverse base pairs fluctuations. In fact, the disruption of a hydrogen bond is more likely to occur whenever a base is flipped out of the stack. Therefore a correlation should exist between the size of the anharmonic parameter and the integral cutoff. 

To pursue this idea I model the DNA chain by the PB ladder Hamiltonian which, beyond being computationally easier to handle than the 3D helical model, depends on the same potential parameters and radial cutoff.

The computational method is based on the path integral formalism which conceives the base pair vibrations as time-dependent paths fluctuating around the bare helix diameter. Focusing on a specific base pair of the chain, in thermal equilibrium with the rest of the DNA fragment, one can pin the initial position for the base pair trajectory and define the initial time probability as the boundary condition. Next, for a given set of model parameters, I calculate the probability for the base pair path to return to the starting spatial position as a function of time. The procedure is reiterated for different choices of the integral cutoff until a good value is found which permits to get a probability plot fitting the boundary condition. In the case the good cutoff value had not to exist, the set of input parameters would be discarded.
Thus, the computed value for the initial time probability provides a benchmark for the evaluation of the set of model parameters and the associated integral cutoff. 

The Hamiltonian model is outlined in Section II, while the first-passage formalism is defined in Section III. The results are presented in Section IV and some final remarks are made in Section V.

\section*{II. Model }

\begin{figure}
\includegraphics[height=8.0cm,width=8.0cm,angle=-90]{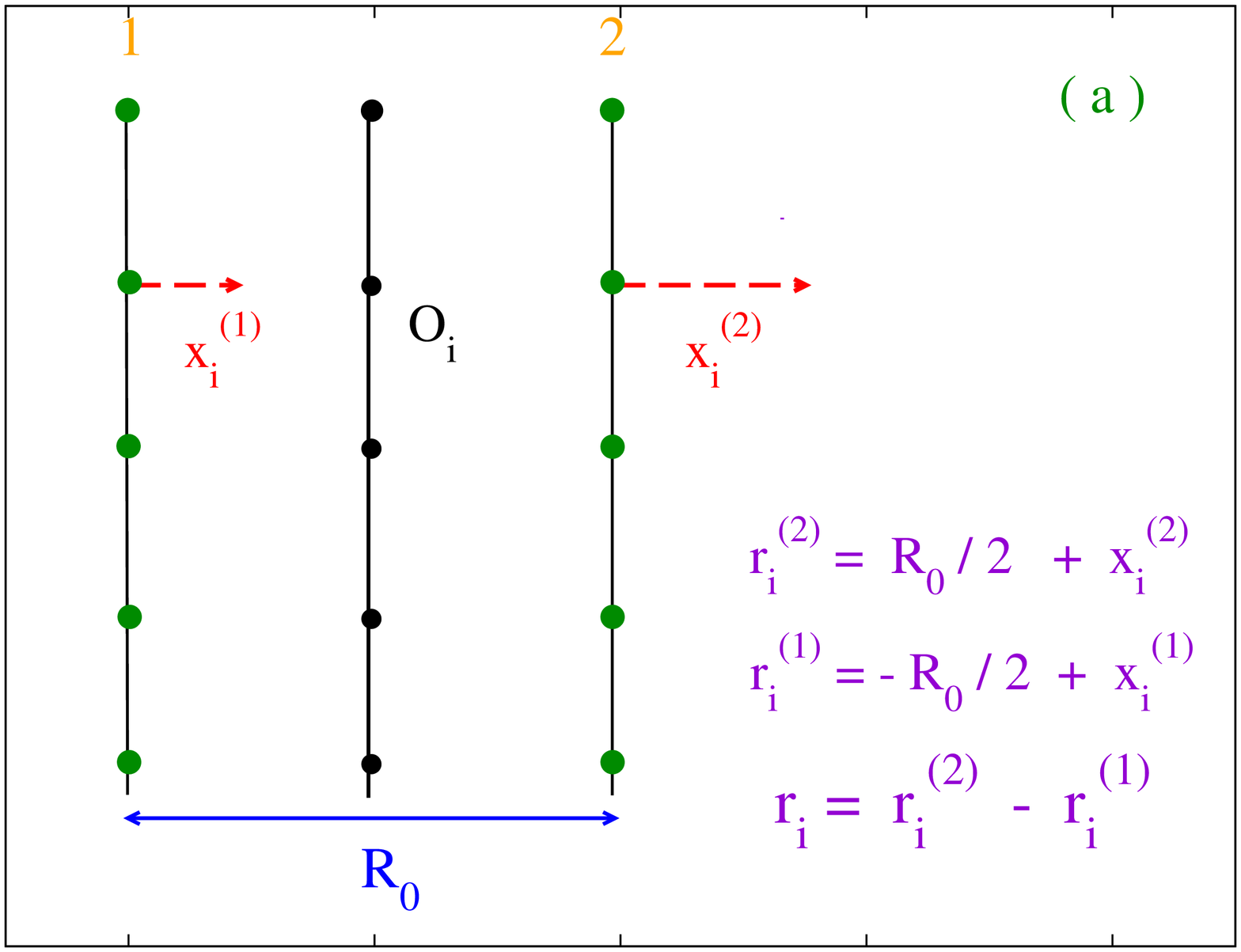}
\includegraphics[height=8.0cm,width=8.0cm,angle=-90]{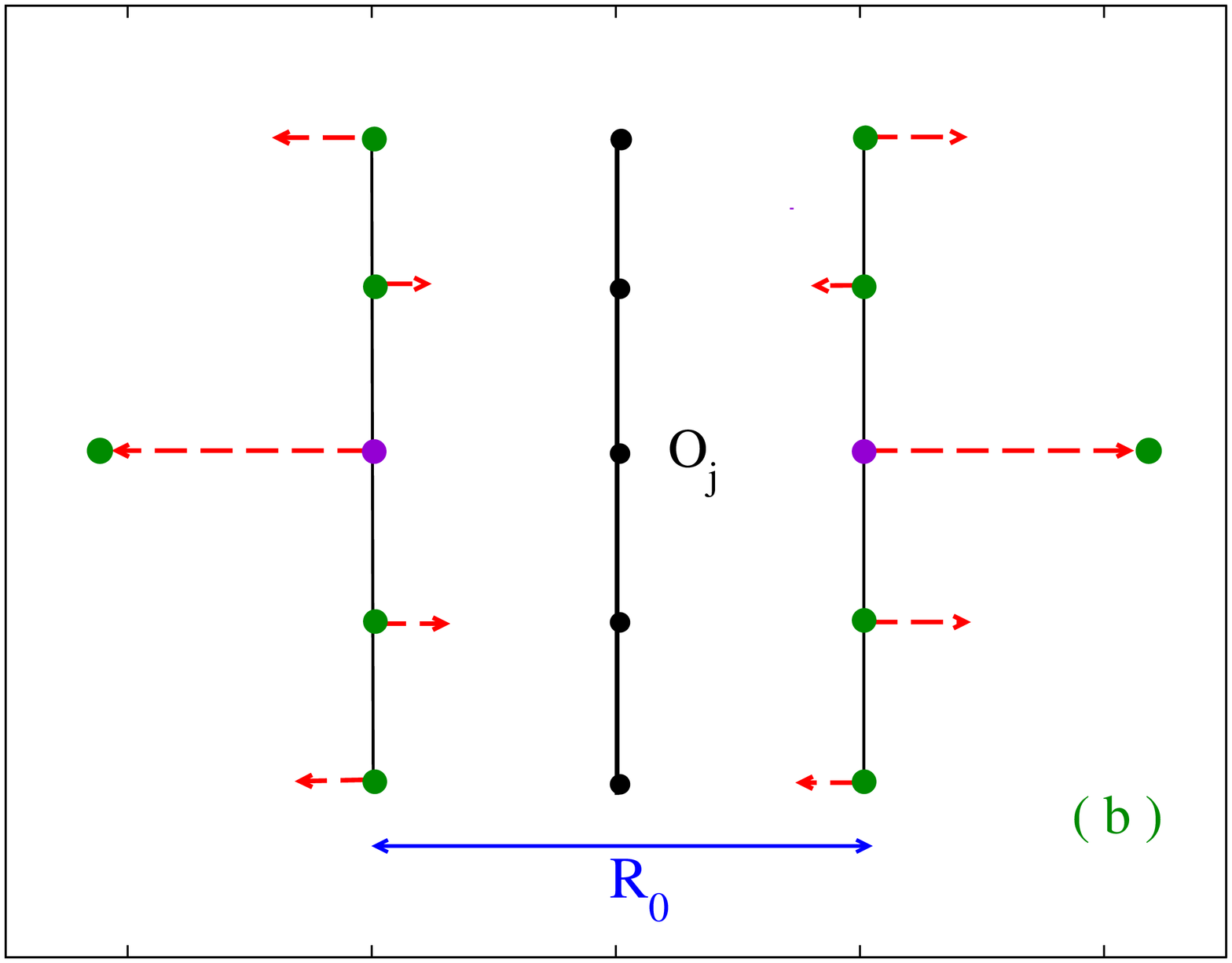}
\caption{\label{fig:1}(Color online)  
Ladder model for an open end chain with $N$ point-like base pairs arranged along the two complementary strands. (a) The transverse base fluctuations, $r_i^{(1,2)}$, are measured with respect to the mid-chain axis hence, the distance $r_i$ between the pair mates is defined with respect to $R_0$. (b) The first-passage probability is computed for the mid-chain $j-th$ base pair which undergoes time dependent fluctuations while maintaining its stacked position.
}
\end{figure}

We begin with the standard model \cite{pey93} for a chain of $N$ point-like base pairs (of reduced mass  $\mu$)  which are arranged like beads along two parallel strands, see Fig.~\ref{fig:1}(a), set at the distance $R_0$. The latter is the bare helix diameter in the absence of fluctuations. For each pair mate, only transverse fluctuations are considered and the only degree of freedom is $r_i$, that is the relative distance between complementary pair mates measured  with respect to $R_0$. The Hamiltonian for this 1D Ladder Model contains the essential contributions to the chain stability due to the base pairing and stacking. The explicit form reads:

\begin{eqnarray}
& &H_{LM} =\, H_a[r_1] + \sum_{i=2}^{N} H_b[r_i, r_{i-1}] \, , \nonumber
\\
& &H_a[r_1] =\, \frac{\mu}{2} \dot{r}_1^2 + V_{1}[r_1] \, , \nonumber
\\
& &H_b[r_i, r_{i-1}]= \,  \frac{\mu}{2} \dot{r}_i^2 + V_{1}[r_i] + V_{2}[ r_i, r_{i-1}]  \, \, , \nonumber
\\ 
& &V_{1}[r_i]=\, D_i \bigl[\exp(-b_i (|r_i| - R_0)) - 1 \bigr]^2  \, , \nonumber
\\
& &V_{2}[ r_i, r_{i-1}]=\, K_{i, i-1} \cdot \bigl(1 + G_{i, i-1}\bigr) \cdot \bigl( |r_i| - |r_{i-1}|\bigr)^2  \, , \nonumber
\\
& &G_{i, i-1}= \, \rho_{i, i-1}\exp\bigl[-\alpha_{i, i-1}(|r_i| + |r_{i-1}| - 2R_0)\bigr]  \, . 
\label{eq:01}
\end{eqnarray}

$H_a[r_1]$ is taken out of the sum as the first base pair lacks the preceding neighbor along the stack.  The Morse potential $V_{1}[r_i]$ models the base pair hydrogen bonds and, due to its hard core, also accounts for the repulsive electrostatic interaction between phosphate groups on complementary strands. 
$D_i$ is the base pair dissociation energy and $b_i$ sets the potential range. While fluctuations may bring the pair mates at a distance even shorter than $R_0$, too negative base pair contractions are energetically discouraged by the repulsive potential core. This physical constraint is implemented in the calculations by retaining only fluctuations whose energy is such that the condition \, $V_{1}[r_i] \leq D_i$ \, is fulfilled, i.e.,  $|r_i| - R_0 \geq - \ln 2/b_i$. 

The stacking interactions are described by the two-particle potential $V_{2}[ r_i, r_{i-1}]$ which depends on the elastic force constant $K_{i, i-1}$ and anharmonic parameters $\rho_{i, i-1}$, $\alpha_{i, i-1}$. The specific form for the nonlinear term $G_{i, i-1}$ has been chosen to model the cooperative effects associated to the bubble formation in the denaturation regime. If the inequality \, $|r_i| - R_0 \gg \alpha_{i, i-1}^{-1}$ \, holds, one mate (or both) of the $i-th$ base pair flips out of the stack thus causing a reduction in the stacking interaction, from  $ \sim K_{i, i-1}(1 + \rho_{i, i-1})$ to $ \sim  K_{i, i-1}$. This energetic gain, weighed by $\rho_{i, i-1}$, favors in turn the disruption of the adjacent base pair  and promotes the formation of fluctuational openings. 
Hence, the anharmonic PB model  accounts for the melting cooperativity in DNA.
As covalent bonds along the stack are stronger than base pair hydrogen bonds, a large amplitude transverse fluctuation sampling the Morse plateau, with energy $D_i$, may not suffice to unstack the base pair. Accordingly the condition $\alpha_{i, i-1} < b_i$ is consistently assumed in the calculations.

While, in principle, the sequence heterogeneity can be introduced in the model via the complete set of five potential parameters, most investigations take only the Morse parameters as base specific (e.g. \cite{campa98}) or extend the heterogeneity effects solely to the harmonic stacking (e.g. \cite{bish12}). Instead, heterogeneous anharmonic parameters have been considered in 3D Hamiltonian models for DNA minicircles \cite{io14a} and in the calculation of the end-to-end distance for sequences in confined environments \cite{io20}.

As discussed in the Introduction, the  Hamiltonian in Eq.~(\ref{eq:01}) (and its harmonic variant with $\rho_{i, i-1}=\,0$) has been applied both to long ($\sim$ kilo base pairs) and short ($\sim 10$ base pairs) chains with parameter values generally adjusted to reproduce the experimentally available melting temperatures.
In short chains, finite size effects which broaden the denaturation transition  can be accounted for by taking open boundary conditions \cite{joy07}. Also the fraying at the chain ends can be incorporated in Eq.~(\ref{eq:01}) by a specific parametrization for the terminal base pairs \cite{weber15,zgarb14}.

\section*{III. Path integral for the first-passage probability}

DNA in solution is constantly subjected to thermal fluctuations which deform the molecular bonds causing transient openings along the chain and associated breathing dynamics. Due to the thermal buffeting,  the base pairs vibrations are an example of constrained one dimensional Brownian motion which can be treated by the path integral formalism \cite{fehi}. Take for instance the mid-chain $j-$ base pair in Fig.~\ref{fig:1}(b) and assume that, at the initial time, the average distance between the pair mates is, $< r_j > =\,R_0$. At any successive time $t$, fluctuations may cause $r_j$ to contract or expand with respect to $R_0$. Accordingly, we define $P_j(R_0,\, t)$ as the probability  that $r_j$ does not return to the initial value up to $t$ and $F_j(R_0,\, t)=\, - d P_j(R_0,\, t) dt$ as the probability that the path will first return to the origin between $t$ and $t + dt$.

To carry out the calculation, we think of any $r_i$ in Eq.~(\ref{eq:01}) as a trajectory $r_i(\tau)$ where $\tau \in [0, \beta]$  is the Euclidean time of the finite temperature path integration, with $\beta$ being the inverse temperature. Accordingly $r_i(\tau)$ can be expanded in Fourier series around  $R_0$:

\begin{eqnarray}
r_i(\tau)=\, R_0 + \sum_{m=1}^{\infty}\Bigl[(a_m)_i \cos\bigl( \frac{2 m \pi}{\beta} \tau \bigr) + (b_m)_i \sin\bigl(\frac{2 m \pi}{\beta} \tau \bigr) \Bigr] \, ,
\label{eq:02}
\end{eqnarray}

with the Fourier coefficients weighing the fluctuational effects \cite{io16b}.

From Eq.~(\ref{eq:02}), it is noticed that the closure condition, $r_i(0)=\,r_i(\beta)$,  is equivalent to the periodic boundary conditions (PBC) generally used to derive the partition function for the system in Eq.~(\ref{eq:01}) by  Transfer Integral methods. The latter usually apply the PBC in the real space by closing the chain into a loop, an assumption which may not be appropriate to deal with short sequences in which the boundary effects matter. This shortcoming is overcome in the path integral method as the closure condition holds on the time axis while the chain is assumed with open ends in real space.  
 
For a free base pair $P_1(R_0,\, t)$ could be calculated as a path integral of the Boltzmann factor for the action associated to $H_a[r_1]$ in Eq.~(\ref{eq:01}). For the $j-th$ base pair interacting with its first neighbors along the stack (Fig.~\ref{fig:1}(b)), we write the general expression for $P_j(R_0,\, t)$ as:

\begin{eqnarray}
& &P_j(R_0,\, t)=\, \int_{r_j(0)}^{r_j(t)} Dr_{j} \exp \bigl[- A_a[r_j] \bigr] \cdot \oint Dr_{1} \exp \bigl[- A_a[r_1] \bigr] 
\cdot \,  \nonumber 
\\
& & \prod_{i=2}^{N}{}^{'} \oint Dr_{i}  \exp \bigl[- A_b [r_i, r_{i-1}] \bigr] \cdot \prod_{\tau=\,0}^{t}\Theta\bigl[r_j(\tau) - R_0\bigr] \, , \nonumber 
\\
& &A_a[r_j]=\,\int_{0}^{t} d\tau H_a[r_j(\tau)] \, , \nonumber 
\\
& &A_a[r_1]=\,\int_{0}^{\beta} d\tau H_a[r_1(\tau)] \, , \nonumber 
\\
& &A_b [r_i, r_{i-1}]=\, \int_{0}^{\beta} d\tau H_b[r_i(\tau), r_{i-1}(\tau)]
\label{eq:03}
\end{eqnarray}

The prime symbol in the product over the base pair index denotes that, for $i=\,j$, $H_b[r_i(\tau), r_{i-1}(\tau)]$  reduces to  $V_{2}[ r_j(\tau), r_{j-1}(\tau)]$. 
 It is pointed out that the  $\oint {D}r_{i}$ and $\int Dr_{j}$ integrations in Eq.~(\ref{eq:03})  are not independent as $r_j$ is coupled to the neighboring base pairs via the two particle stacking potential in $A_b [r_i, r_{i-1}]$. For instance, taking a fragment with $N=\,5$ as in Fig.~\ref{fig:1}(b),  the fluctuation of the mid-chain base pair \, $r_j$ \, is coupled to the \, $i=2,4$\, integral contributions in $\Pi^{'}$.

For the $N-1$ paths in the stack (the $j-th$ path is treated separately), $\oint {D}r_i$ is the measure of integration over the Fourier coefficients defined by:

\begin{eqnarray}
& &\oint {D}r_{i} \equiv  \prod_{m=1}^{\infty}\Bigl( \frac{m \pi}{\lambda_{cl}} \Bigr)^2  \times \int_{-\Lambda(T)}^{\Lambda(T)} d(a_m)_i \int_{-\Lambda(T)}^{\Lambda(T)} d(b_m)_i \, , \, 
\label{eq:04}
\end{eqnarray}

where  $\lambda_{cl}$ is the classical thermal wavelength and $\Lambda(T)$ is the temperature dependent cutoff which truncates the configuration space for the base pair fluctuations. As the functional measure $\oint {D}r_{i}$ normalizes the kinetic action 

\begin{eqnarray}
\oint {D}r_i \exp\Bigl[- \int_0^\beta d\tau {\mu \over 2}\dot{r}_i(\tau)^2  \Bigr] = \,1 \, ,
\label{eq:05} \,
\end{eqnarray}

we have a condition intrinsic to the path integral method \cite{io05,io09} to establish the cutoff $\Lambda(T)$ which controls the temperature dependent range of the base pair fluctuations. Using  Eqs.~(\ref{eq:02}),~(\ref{eq:04}), the l.h.s. of Eq.~(\ref{eq:05}) transform into a product of independent Gau{\ss}ian integrals whose solution yields:
$\Lambda(T)=\,{{U \lambda_{cl}} / {m \pi^{3/2}}}$, with  $U$ being a dimensionless parameter which is set numerically, i.e. $U=\,2$ \cite{n1}.

Incidentally, it is worth noticing that $A_a[r_j]$ in the second of Eqs.~(\ref{eq:03}) has the form of a random variable for which one can in principle calculate the probability distribution  \cite{maj05}. In this specific problem however the Fourier coefficients for the path $r_j(\tau)$ and the Morse parameters (which set the value of $A_a[r_j]$) cannot be varied independently, their correlation arising from the physical constraint mentioned below Eq.~(\ref{eq:01}).

\subsection*{A. Numerical procedure}

For the $j-th$ base pair, we still write the path as in Eq.~(\ref{eq:02})  but the Heaviside function $\Theta[..]$ in the first of Eq.~(\ref{eq:03}) enforces the condition that $r_j(\tau)$ does not return to $R_0$ for any $\tau \in [0, t]$. 

Accordingly, for a given $t$, the numerical programme checks, at any \, $\tau$ in the range, the amplitude of the fluctuation $r_j(\tau)$ and discards those sets of coefficients which don't comply with the constraint imposed by the $\Theta[..]$. Further, note that Eq.~(\ref{eq:05}) does not hold for $r_j(\tau)$  as this trajectory, being defined up to $r_j(t)$, is not closed for any $t < \beta$. It follows that, for the $j-th$ base pair,  the  integration over the Fourier coefficients, defined by $\int_{r_j(0)}^{r_j(t)} Dr_{j}$, is truncated by a cutoff \, $\Lambda(T)_{j}=\,{{U_j \lambda_{cl}} / {m \pi^{3/2}}}$, with tunable $U_j$. While, in general, $U_j$  may differ from $U$, the two cutoffs should however be of the same order of magnitude as the fluctuations of neighboring base pairs are expected to have similar maximum amplitudes.
Thus, the first of Eq.~(\ref{eq:03}) is computed by increasing the number of paths in the integrand until numerical conv­ergence is achieved, each path being generated by a set of Fourier coefficients. The presented results are obtained by taking, at any time value,  $\sim 2.8 \cdot 10^{6}$ paths for each dimer in the chain.

Let's focus now on $P_j(R_0,\, t)$ at the initial time to extract the fundamental criterion for our analysis. From Eq.~(\ref{eq:02}), one gets \, $r_j(0)=\, R_0 + \sum_{m=1}^{\infty}(a_m)_j$. As the Fourier coefficients are integrated on an even domain, the initial probability $P_j(R_0,\, 0)$ to have the $j-th$ fluctuation larger than $R_0$ should be $ \sim 1/2$. The approximation sign originates from the fact that, for specific values of the Morse potential width, too negative fluctuations \, $|r_j(0)| - R_0$ \, are not included in the computation as discussed after Eq.~(\ref{eq:01}). In these cases, $P_j(R_0,\, 0)$ gets somewhat larger than $1/2$.
Thus,  $ P_j(R_0,\, 0) \sim 1/2$ \, is the general condition to be fulfilled by the model Hamiltonian for any choice of the potential parameters and integration cutoff.

\section*{IV. Results }

The formalism is applied to a homogeneous fragment of $N=\,5$ base pairs testing five sets of parameters used in different studies for the Hamiltonian in Eq.~(\ref{eq:01}). Such studies are carried out for chains of various length. However, for the present purpose of establishing a relation between integral cutoff and model parameters, the length of the fragment is irrelevant as: \textit{i)} the cutoff truncates the radial fluctuations of a single base pair, \textit{ii)} the Morse parameters model the inter-strand force for a single base pair, \textit{iii)} the stacking parameters refer to first neighbors interactions along the stack.

For those works, i.e. refs.\cite{campa98}, \cite{the10}, dealing with heterogeneous sequences, the chosen parameters are those assumed for AT base pairs. The sets are listed in Table~\ref{table:1}, the short notation: \, $K_{i}\equiv \,K_{i,i-1}$, $\rho_{i}\equiv \, \rho_{i,i-1}$, $\alpha_{i}\equiv \,\alpha_{i,i-1}$, being consistent with the homogeneity of the sequence. The parameters hold here for all base pairs in the fragment. 
While the sets in the first three rows are substantially similar mostly for the choice of the anharmonic parameters, a much larger $\rho_{i}$ value is assumed in the last two sets. Also note that, in ref.\cite{the10}, the sizeable $\rho_{i}$ is joined by an anomalously low elastic constant, a factor forty smaller than the lower bound in the range of the experimentally reported force constants \cite{eijck11}.

\begin{table}
\begin{center}
 \begin{tabular}{|c|  c|  c|  c|  c|  c| } 
 \hline
  & $D_{i}\,$ &  $b_{i}\,$ &  $K_{i}\,$ & $\rho_{i}\,$ & $\alpha_{i}\,$ \\ [0.5ex] 
 \hline\hline
 (Zh97) & 38 & 4.45 & 42  & 0.5 & 0.35  \\ 
 \hline
 (CG98) & 50 & 4.2 & 25  & 2.0 & 0.35  \\
 \hline
 (Zo09) & 30 & 4.2 & 60  & 1.0 & 0.35  \\  
 \hline
 (Th10) & 125 & 4.2  & 0.45 & 50 & 0.2 \\
 \hline
 (SS17) & 50 & 4.2 & 10 & 50 & 0.35 \\ [1ex] 
 \hline
\end{tabular}
\end{center}
\caption{Sets of parameters for the ladder model in Eq.~(\ref{eq:01}) assumed by various Authors.  (Zh97) denotes ref. \cite{zhang97}.  (CG98) is ref. \cite{campa98}. (Zo09) is ref. \cite{io09}. (Th10) is ref. \cite{the10}. (SS17) is ref. \cite{singh17}. These five sets have been used in Figs.~\ref{fig:2} - ~\ref{fig:5}, respectively to compute $P_j(R_0,\, t)$ in Eq.~(\ref{eq:03}). 
$D_{i}$'s are in units meV. $b_{i}$ are in $\AA^{-1}$. $K_{i}$ are in $meV \cdot \AA^{-2}$. $\rho_{i}$ are dimensionless. $\alpha_{i}$ are in $\AA^{-1}$. }
\label{table:1}
\end{table}

\begin{figure}
\includegraphics[height=8.0cm,width=8.0cm,angle=-90]{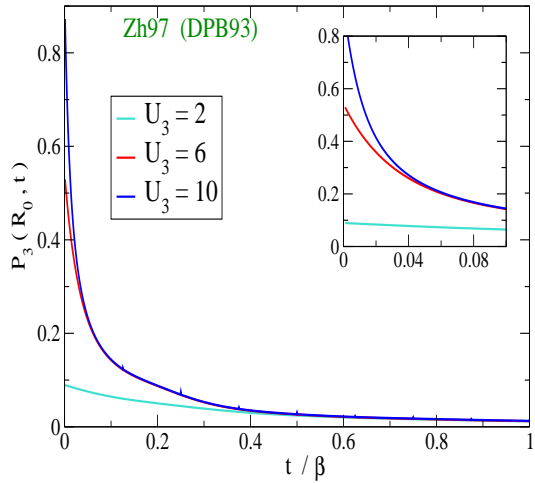}
\caption{\label{fig:2}(Color online)  Probability that the fluctuation amplitude for the mid-chain base pair (in Fig.~\ref{fig:1}(b))  remains larger than $R_0$ up to $t$. Following Eq.~(\ref{eq:02}), the time is measured on the inverse energy scale. The calculation in Eq.~(\ref{eq:03}) is carried out assuming the model parameters of ref. \cite{zhang97} which, in turn, are similar to those of ref. \cite{pey93}.
Three values are chosen for the tunable cutoff on the Fourier coefficients integration. The inset magnifies the probability function close to the time origin.
}
\end{figure}

First, Eq.~(\ref{eq:03}) is computed at room temperature assuming the model parameters in ref. \cite{zhang97}, which are close to those used in the DPB Hamiltonian of ref. \cite{pey93}. The calculation is performed  by tuning $U_j$. The plots for the mid-chain base pair probability versus time are shown in Fig.~\ref{fig:2}, for three values of the integral cutoff \, $U_{j=3}$. It is found that for $U_{3}=\,6$, the curve correctly fits the condition \, $ P_j(R_0,\, 0) \sim 1/2$ as better noticed in the inset. To be precise, the initial probability is evaluated at $t / \beta =\,10^{-3}$ as the range \,$[0, \beta]$ is partitioned in $1000$ points. The general trend suggested by these plots is that the computed $ P_3(R_0,\, t)$ is markedly higher, mostly in the low $t$ region, if larger cutoffs are assumed.  

\begin{figure}
\includegraphics[height=8.0cm,width=8.0cm,angle=-90]{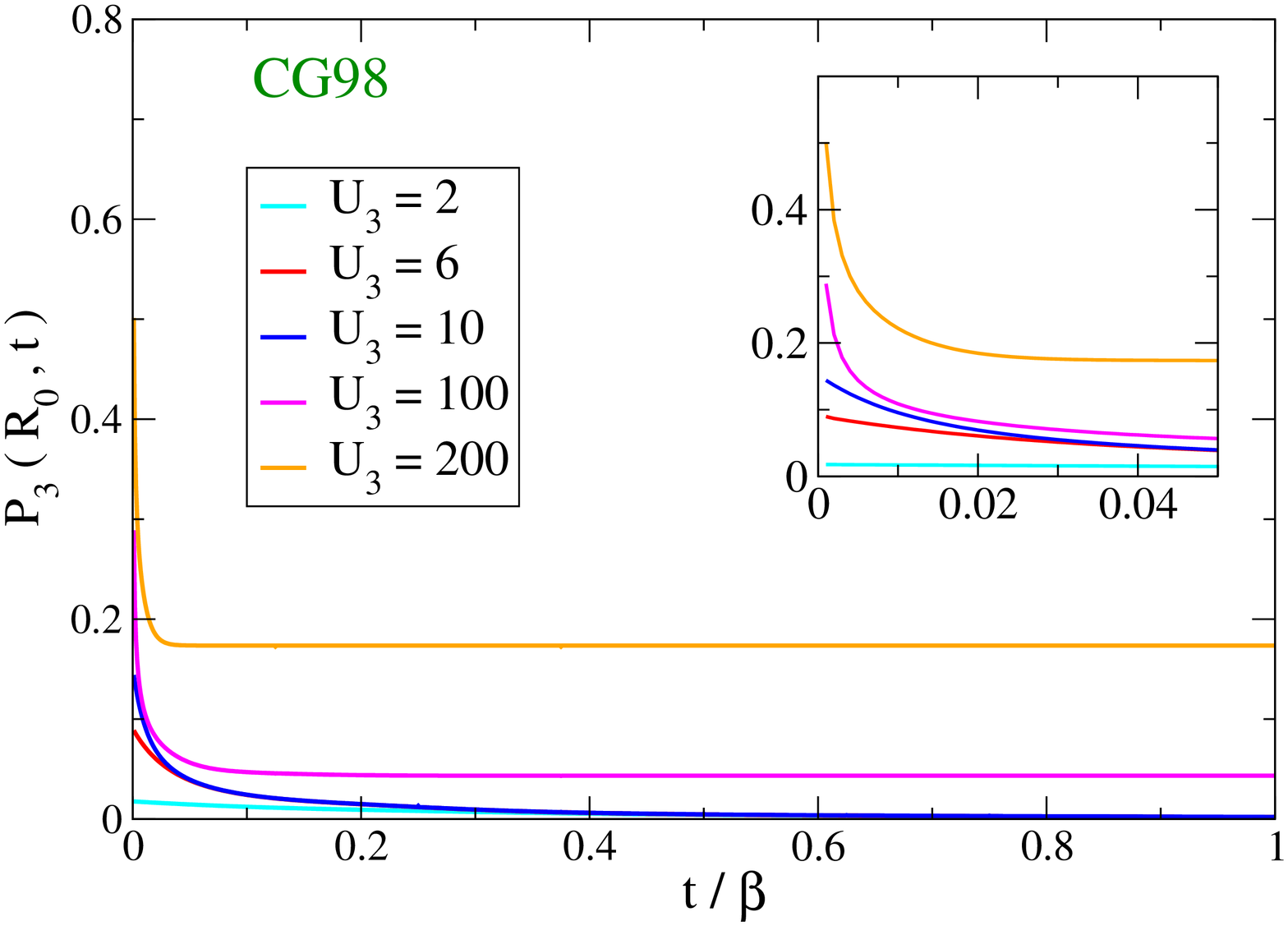}
\caption{\label{fig:3}(Color online)  As in Fig.~\ref{fig:2} but for the model parameters of ref. \cite{campa98}.
}
\end{figure}

Next, let's consider the parameter set of ref.\cite{campa98} whose $\rho_{i}$ is a factor four larger than in ref. \cite{zhang97} whereas $K_{i}$ is taken smaller. In this case, see Fig.~\ref{fig:3}, the initial probability still attains the $1/2$ value but this occurs for a high cutoff, i.e. $U_{3}=\,200$, which is two orders of magnitude larger than the value set by the condition in Eq.~(\ref{eq:05}) for the other four base pairs in the fragment \cite{n1}. 

\begin{figure}
\includegraphics[height=8.0cm,width=8.0cm,angle=-90]{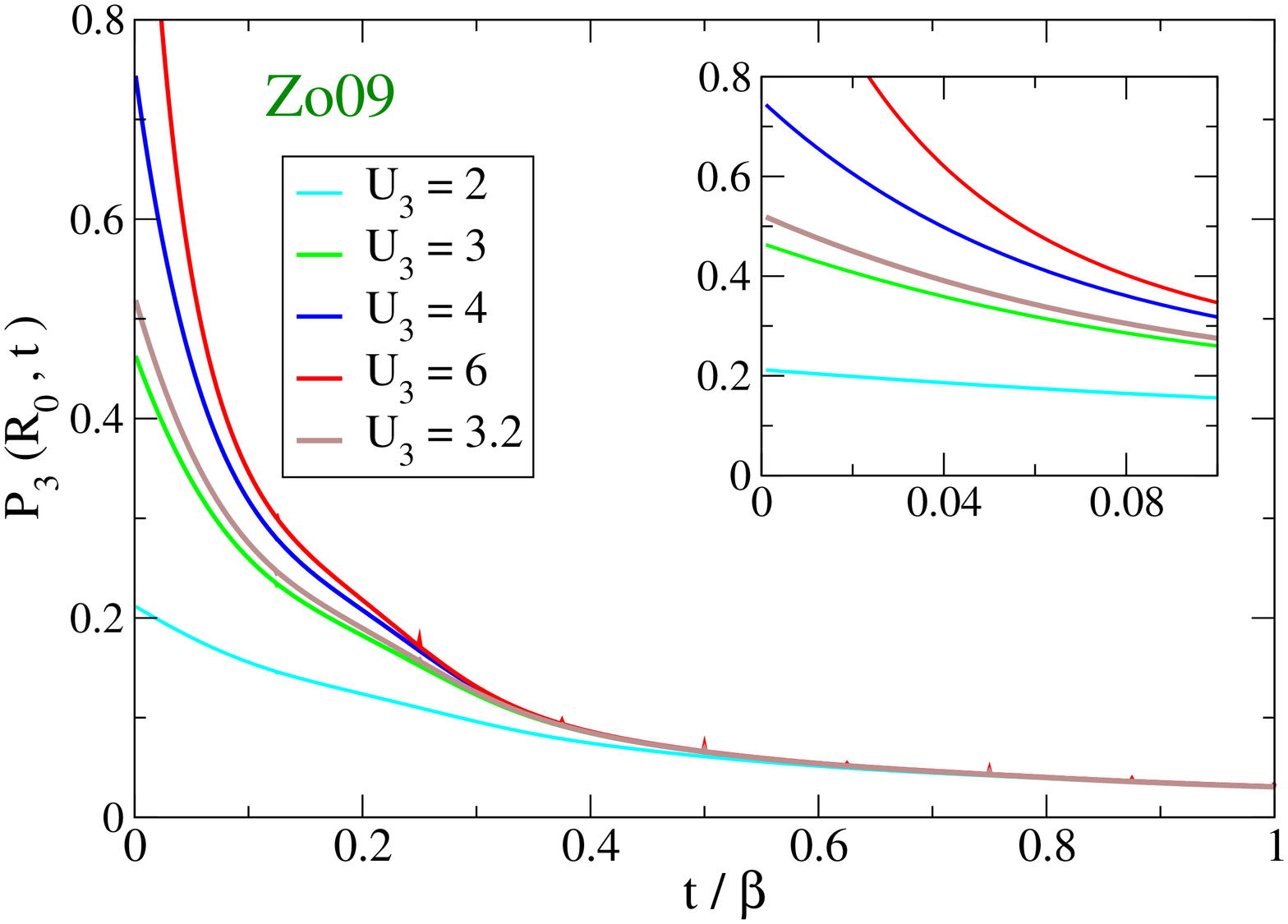}
\caption{\label{fig:4}(Color online)  As in Fig.~\ref{fig:2} but for the model parameters of ref. \cite{io09}.
}
\end{figure}

Taking the third set in Table~\ref{table:1}, it is seen in Fig.~\ref{fig:4} that $P_3(R_0,\, t)$ has a pronounced sensitivity to the cutoff value in the low $t$ range. This feature,  similar to the results of Fig.~\ref{fig:2}, also stems from the relatively low base pair dissociation energies chosen in both studies. The condition \, $ P_3(R_0,\, 0) \sim 1/2$ is verified by choosing \,$U_{3}=\,3.2$. Given this value, the cutoff for the first Fourier component in the path of Eq.~(\ref{eq:02}) is \, $\Lambda_{T} \sim \,0.7 \, \AA$ hence, the path fluctuation amplitude around $R_0$ is $\sim 1 \,\AA$. This appears as a reasonable estimate for the room temperature base pair fluctuations \cite{n2}.
  
\begin{figure}
\includegraphics[height=8.0cm,width=8.0cm,angle=-90]{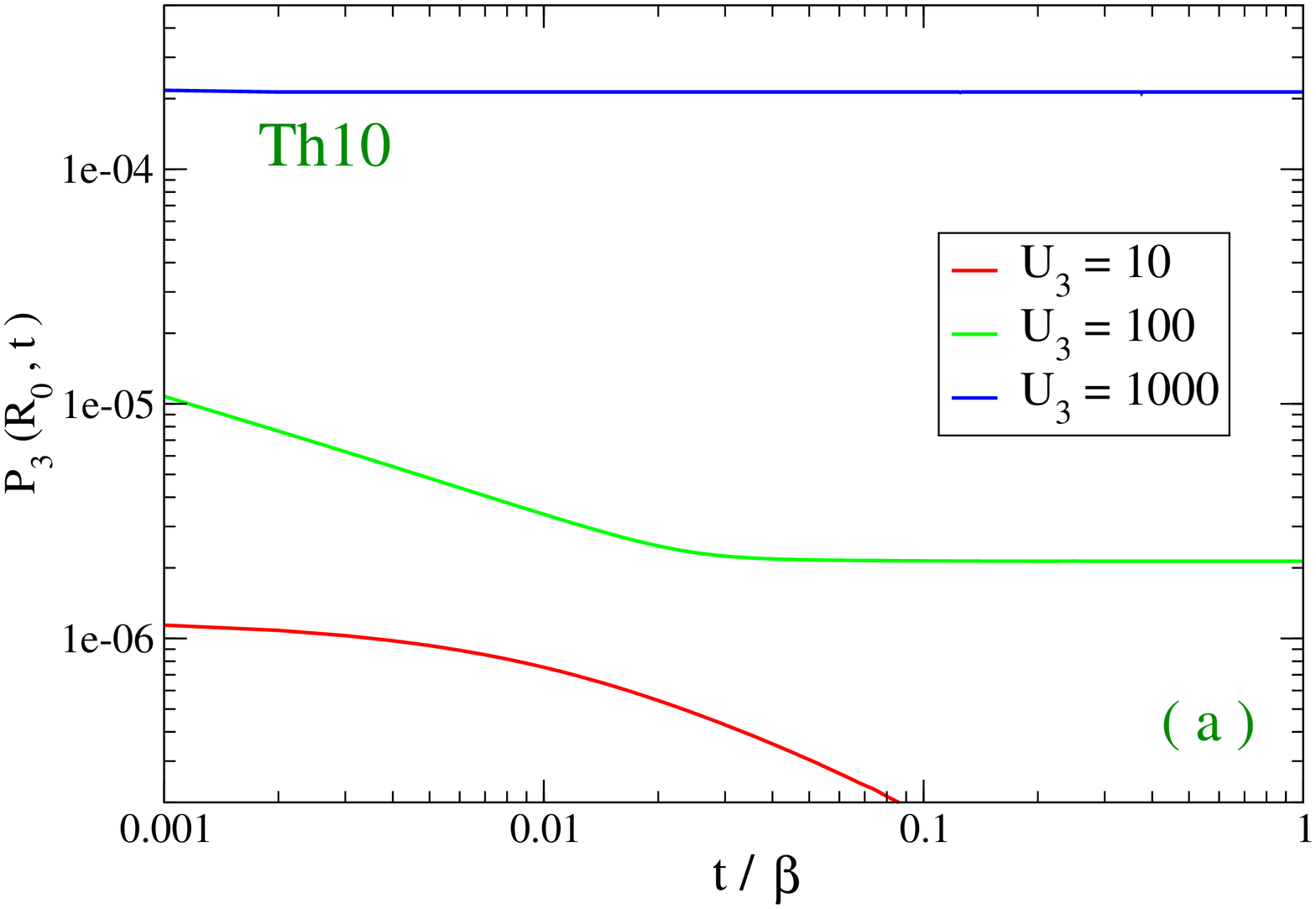}
\includegraphics[height=8.0cm,width=8.0cm,angle=-90]{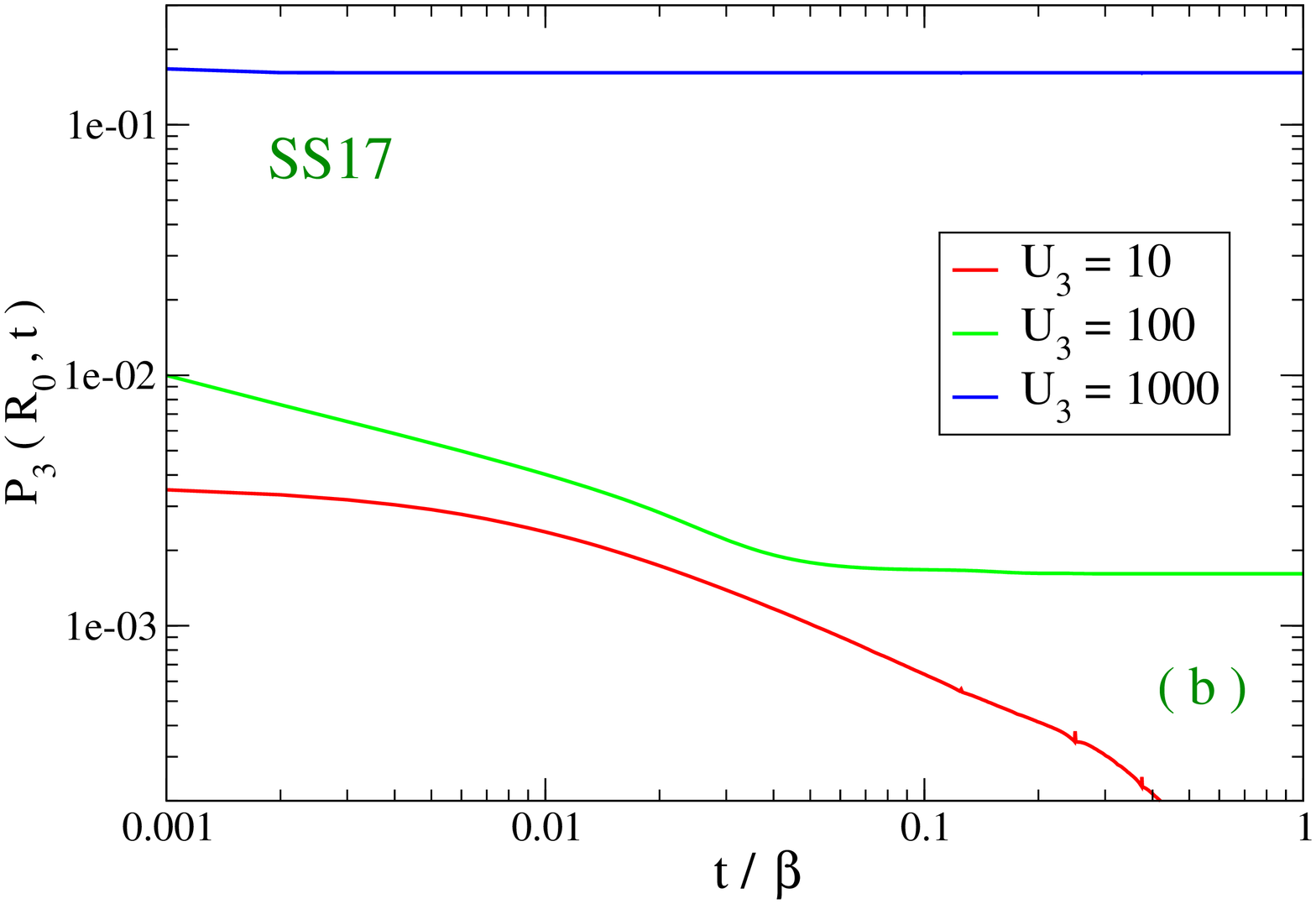}
\caption{\label{fig:5}(Color online)  As in Fig.~\ref{fig:2} but for the model parameters of : (a) ref. \cite{the10}, (b) ref. \cite{singh17}.  A log-log plot is used in both panels.
}
\end{figure}

The plots obtained by assuming the fourth and fifth set of parameters in Table~\ref{table:1} are shown in Fig.~\ref{fig:5}. Neither set permits to fulfill the initial condition as the computed probabilities are always too small, however high one may choose the cutoff $U_{3}$. The poor performance of the last two  sets is clearly ascribed to the anomalously high nonlinear parameter $\rho_{i}$ while the very high dissociation energy chosen in ref. \cite{the10} also contributes to squeeze the probability towards vanishing values.

Summing up, the results obtained in Figs.~\ref{fig:2} -~\ref{fig:4} indicate that the non-linear stacking $\rho_{i}$ should be taken in the range $\sim [0.5 - 2]$ whereby  possible variations should be ascribed to the choice of the other potential parameters. This estimate appears also in line with the original assumptions made for the DPB model \cite{pey93}.

Finally, we focus on the first three parameter sets which have passed the test of the initial probability and wonder to which extent the specific cutoffs selected in Figs.~\ref{fig:2} -~\ref{fig:4} may be further discerned on the base of some physical property. For instance it is interesting to consider how the lifetime of the path fluctuation may be affected by the cutoff size.   

\begin{figure}
\includegraphics[height=8.0cm,width=8.0cm,angle=-90]{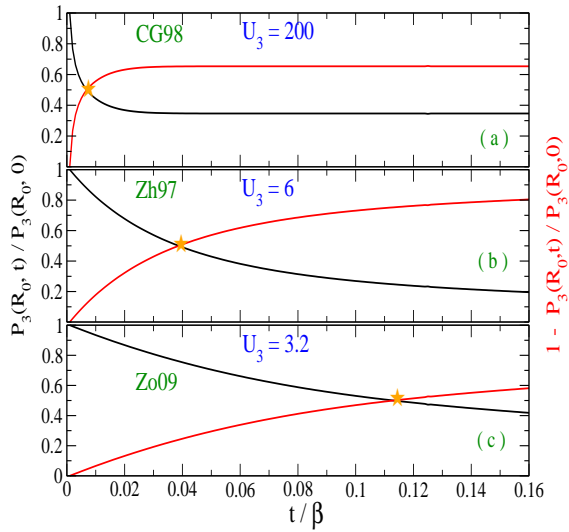}
\caption{\label{fig:6}(Color online)   Normalized probabilities versus time, for the best cutoff values  derived respectively from: (a) Fig.~\ref{fig:3}, (b) Fig.~\ref{fig:2}, (c) Fig.~\ref{fig:4}.  The orange $\bigstar$ mark the times at which the normalized probabilities are halved with respect to their initial values.
}
\end{figure}

To this purpose,   Eq.~(\ref{eq:03}) is computed for the first three sets in Table~\ref{table:1} by keeping only the paths which fulfill the Heaviside function constraint. Accordingly, the normalization condition \, $ P_3(R_0,\, 0) =\, 1$ \, holds if the calculation is carried out for $U_3=\,6, \, 200, \, 3.2$, respectively. Next, we define $t^{*}$ as the lifetime of a path fluctuation larger than $R_0$ and estimate $t^{*}$ as the time at which \,$ P_3(R_0,\, t)$ drops to $1/2$. The results are displayed in Fig.~\ref{fig:6}. It is found that \, $t^{*}$  gets the shortest value for the parameter set which requires the largest $U_3$  as broad fluctuations may quickly bring back the path to the $R_0$ threshold. Consistently, by taking parameter sets with decreasing $U_3$ values, $t^{*}$ becomes longer as smaller amplitude fluctuations take a longer time to restore the initial condition.

It is also noticed that the formalism here developed may be adapted to evaluate the lifetime of an open base pair which triggers the formation of a fluctuational bubble \cite{gueron,russu,bonnet03,bandyo11,hess17} also in the presence of external periodic forces \cite{bandyo19}. This could be done by modifying the threshold in the argument of $\Theta[..]$ in Eq.~(\ref{eq:03}) thus assuming that the base pair paths remain larger than $R_0 + \delta$ up to a given $t$, with $\delta$ typically of order $2 \AA$ \cite{io11a,io11b}. This calculation will however need a hydrogen bond potential which incorporates a barrier for base pair re-closing \cite{pey09a,singh09} and a description of the double helix structure accounting for the bending and twisting of the strands \cite{io18c}, that is more complex than that proposed in Eq.~(\ref{eq:01}).

\section*{V. Conclusions }

Coarse-grained Hamiltonian models generally provide a convenient description of biomolecules properties but their predictive capability is often hindered  by the uncertainties regarding the choice of some model parameters. I have focused here on a well-known one dimensional model, depicting the DNA molecule as made of two parallel strands, proposed long ago to account for the melting transition. While, in some previous works, I had pointed out the shortcomings of this model as for the calculation of the flexibility properties of the 3D helical molecule,  it is recognized that the 1D ladder model offers a computationally useful representation of the main forces which stabilize DNA in terms of a set of input parameters.  Such parameters can be related to experimental quantities when available for specific sequences and, further, can be used in more realistic mesoscopic models which account for the DNA twisting and bending fluctuations.  In particular, while substantial convergence has been achieved as for the parametrization of the hydrogen bonds for the inter-strand base pair interactions, significant discrepancies are found in the literature among the values of the stacking parameters which govern the intra-strand forces, in particular their non-linear components. These differences may have a considerable impact on the estimates of DNA flexibility properties such as persistence length and looping probability. Motivated by this observation, I have proposed a novel approach which applies the path integral formalism to derive the first-passage properties of the 1D Hamiltonian model and computed the time dependent probability, for a specific base pair in the chain, to keep a relative distance between the mates larger than a given threshold, i.e. the bare helix diameter. However large such distance may get, the mid-chain base pair is assumed to be in thermal equilibrium with the other $N-1$ base pairs along the stack.

The idea underlying this approach is that the base pair vibrations are time dependent paths subjected to the constant action of the thermal fluctuations. Calculating the probability in Eq.~(\ref{eq:03}) as a function of time for a short homogeneous fragment, I have selected those sets of input parameters which fulfill a general constraint imposed by the initial conditions. The main results brought about by this investigation are twofold: first, it shows that too large values for the non-linear stacking parameter \,$\rho _i$ are ruled out, at least for a chain in the double-stranded configuration; second, it permits to determine the cutoffs on the spatial range of the base pair fluctuations consistently with the specific set of model parameters thus avoiding those arbitrariness which are known to affect other computational methods generally applied to the DNA Hamiltonian model, e.g. transfer integral techniques or molecular dynamics simulations. A similar procedure can be
used also for heterogeneous sequences by choosing a larger set of model parameters which accounts for the sequence specificities in the one particle and two particles potential.  Finally, it is suggested that the presented formalism can be further developed to estimate the lifetime of a transiently open base pair provided that a three dimensional model for the helical chain, more realistic than the ladder model here considered, is assumed.


\begin{thebibliography}{widest-label}
\bibitem{n1}
Taking the series expansion for the Gau{\ss}ian integral \cite{grad}:

$\int_{0}^{U} dx \exp(-x^{2}) =\, \sum_{0}^{\infty} \frac{(-1)^{k} U^{2k+1}}{k! (2k+1)}$ , it is found that the Poisson value $\sqrt{\pi}/2$ is accurately fit by truncating the series with $k_{max}=\,10$ and setting $U=\,2$.
\bibitem{n2}
The path integral computation of the DNA thermodynamic properties is generally carried out by taking only the first Fourier component in the expansion of Eq.~(\ref{eq:02}). This suffices to get a numerical convergent partition function. Likewise, Eq.~(\ref{eq:03}) is computed by assuming $m=\,1$.
\end{thebibliography}
\end{document}